\documentclass{article}
\usepackage{amsmath,amssymb}
\usepackage[dvips]{graphicx}
\usepackage{cite}

\def\abstract#1{\vskip 7mm 
        \begin{center}{\large Abstract}\par \smallskip
                \begin{minipage}[c]{12cm}
                        \small #1
                \end{minipage}
        \end{center}
}
\def\title#1{\begin{center}{\Large\bf #1}\end{center}}
\def\author#1{\vskip 5mm \begin{center}{#1}\end{center}}
\def\address#1{\begin{center}{\it #1}\end{center}}
\makeatletter
\@ifundefined{lesssim}{}{}
\@ifundefined{gtrsim}{}{}
\def\vereq#1#2{\lower3pt\vbox{\baselineskip1.5pt \lineskip1.5pt
\ialign{$\m@th#1\hfill##\hfil$\crcr#2\crcr\sim\crcr}}}
\makeatother

\newcommand{\dd}{D6-D2$^*$}
\newcommand{\ddp}{D4-D0$^*$}
\newcommand{\Rb}{\mathbb{R}}
\newcommand{\Enh}{Enhan\c{c}on}
\newcommand{\enh}{enhan\c{c}on}
\newcommand{\oo}{$\bigcirc$}
\newcommand{\xx}{$\times$}
\newcommand{\ls}{\ell_s}
\newcommand{\del}{\partial}
\newcommand{\vev}[1]{\langle #1 \rangle}
\newcommand{\nn}{\nonumber}
\begin{document}
%
\begin{flushright}
 KUCP-0194\\
 {\tt gr-qc/0108084}
\end{flushright}

\title{%
{\Enh} and Resolution of Singularity
}
\author{%
  Satoshi Yamaguchi\footnote{E-mail:yamaguch@phys.h.kyoto-u.ac.jp}
}
\address{%
  Graduate School of Human and Environmental Studies, \\
  Kyoto University, Kyoto, 606--8501, Japan
}

\abstract{ We review the {\enh} mechanism proposed by Johnson, Peet and
Polchinski.  If we consider the D6-brane wrapped on K3, then there
appears a naked singularity called ``repulson'' in the supergravity
solution.  But this singularity is resolved by a shell structure called
``{\enh}''. In the interior of the {\enh}, the abelian gauge symmetry is
enhanced to a nonabelian one, and ordinary supergravity is no more
reliable.  We also review the interpretation of {\enh} as fuzzy sphere.
This paper is the contribution to the proceedings of ``Frontier of
Cosmology and Gravitation", April 25-27 2001, YITP.  }

\section{Introduction}

The string theory includes the graviton in its massless spectrum, and
each order of the loop expansion is finite. Therefore, the string theory
is thought as the ``quantum theory of the gravity'' (For a review of the
string theory, see \cite{Joe}).  However, the formulation of the string
theory depends on background, and we can only formulate it as a
fluctuation from the background.

Let us consider what happens when the background has some geometrical
singularity. If one can formulate the string theory consistently on the
geometrically singular background, let us call it ``resolution of
singularity in the string theory''. For a review of singularities in the
string theory, see \cite{Natsuume}.

Orbifolds are the most famous examples of resolution of singularity in
string theory. The string theory on an orbifold can be formulated
consistently by including the twisted sectors. The twisted sectors
correspond to the resolution modes.

Also D-branes are interesting objects from the point of view of
singularities. A D-brane is the object on which open string can
end\cite{Joe}. On the other hand, a D-brane can also be described as a
solution of the classical supergravity\cite{Soda} --- the low energy
effective theory of the string theory.  The supergravity description and
the open string description of a D-brane do not look very similar at
first sight. The evidences of the equivalence of these two are as follows.
\begin{itemize}
 \item The supergravity description and the open string description of a
       D-brane have the same charge called Ramond-Ramond charge.
 \item Both of them are BPS saturated states. BPS saturated states are
       the lowest energy states out of those which have the same charge,
       that is, extremal in terms of the relativity.
\end{itemize}
For this reason, we regard them as the same object.

If we describe this D-brane as a solution of
the supergravity, there often appear naked singularities. But,
D-branes are not ``bad'' objects in open string description,
the singularities in supergravity description should be resolved
by some mechanisms.

The singularity treated in this paper --- repulson singularity --- is
one of the D-brane singularities\cite{Johnson:2000qt}.  It does not
resemble to ordinary D-brane singularities in the supergravity
description. However, we know that the string theory on this D-brane
background is not singular, therefore, the geometric singularity is
resolved.  We intend to see how this singularity is resolved from the
geometrical point of view and clarify what happens near the
singularity. In this point of view, Johnson, Peet and
Polchinski\cite{Johnson:2000qt} propose that there appears a special
radius called ``{\enh} radius'' and D-branes form a ``shell'' at this
radius. And in inner region of this radius, the geometry seems to be
flat and there are no singularity.  We review this ``{\enh} mechanism''
in this paper.

The organization of this paper is as follows. In section \ref{sec-dd},
we consider the D-brane configuration called ``{\dd} system'' and see
how the singularity appears. We also analyze this system by the
D-brane probe and see a special radius called ``{\enh} radius''
appears. In section \ref{sec-gauge} we interpret the {\enh} in terms
of the gauge theory and explain the proposal of the ``shell''. In section
\ref{sec-fuzzy}, we introduce another point of view of the {\enh} shell
as a fuzzy sphere.

\section{{\dd} system}\label{sec-dd}

Let us consider the Type IIA string theory on $\Rb^{5,1}\times$K3 . Here,
the manifold ``K3'' is a 4-dimensional Ricci flat
K\"ahler simply connected compact manifold. The manifolds which satisfy
the above conditions are all diffeomorphic. For a review of string
theory on K3 see \cite{Aspinwall:1996mn}.

We consider $N$ D6-branes wrapped on whole K3 in the above theory.  In
the view of the 6-dimensional theory, these D6-branes look 2-branes.
This system has not only $N$ D6-brane charge but also $(-N)$ D2-brane
charge\cite{Bershadsky:1996sp} of shown in table \ref{D6-D2}. This is
because K3 is curved and the stringy correction arises. Actually, these
negatively charged D2-branes are not the ordinary anti-D2-branes and
have negative contribution to the brane tension. We call this system
``{\dd} system''.  This {\dd} system is BPS saturated and conserves 8
supercharges.

\begin{table}[h]
\begin{center}
  \begin{tabular}{|c|c|c|c|c|c|c||c|c|c|c|}\hline
    &0 &1 &2 &3 &4 &5 &6 &7 &8 &9 \\ \hline
    &\multicolumn{6}{|c||}{$\Rb^{5,1}$} &\multicolumn{4}{|c|}{K3} \\
    \hline D6 &\oo &\xx &\xx &\xx &\oo &\oo &\oo &\oo &\oo &\oo \\
    \hline D2${}^*$ &\oo &\xx &\xx &\xx &\oo &\oo &\xx &\xx &\xx &\xx \\
    \hline
\end{tabular}
\end{center}
 \caption{{\dd} system. The branes spread to the direction
marked by {\oo}, and do not spread to the direction marked by {\xx} .
There are actually no D2-brane, but we write D2-brane
in this table in order to express the D2-brane charge.}
 \label{D6-D2}
\end{table}

\subsection{Classical solution}
The low energy effective theory of the type IIA string theory is the
type IIA supergravity.
The type IIA supergravity includes the following fields as bosonic fields.
\begin{itemize}
 \item NSNS fields --- $g_{MN}$ (metric), $B^{(2)}$ (NSNS 2-form), $\Phi$ (dilaton).
 \item RR fields --- $C^{(1)}$ (RR 1-form), $C^{(3)}$ (RR 3-form).
\end{itemize}
A D6-brane has magnetic charge of the gauge field $C^{(1)}$. It is
convenient to denote the electro-magnetic dual of $C^{(1)}$ by 7-form
$C^{(7)}$. Then, A D6-brane has electric charge of $C^{(7)}$. 
On the other hand, D2-brane
has electric charge of $C^{(3)}$. Sometimes it is convenient to use
5-form $C^{(5)}$, which is the electro-magnetic dual of $C^{(3)}$.

Now we write down the classical solution of {\dd} system.  Let the
longitudinal direction on the brane $\mu,\nu=0,4,5$, and their flat
metric $\eta_{\mu\nu}={\rm diag}(-1,1,1)$. We denote the transverse
direction of the branes as $i,j=1,2,3$.  Let us also define the radial
coordinate $r^2=x^ix^i$. By using these notations, the solution
can be written down as
\begin{align}
 &ds^2=Z_2^{-1/2}Z_{6}^{-1/2} \eta_{\mu\nu}dx^{\mu}dx^{\nu}
+Z_2^{1/2}Z_{6}^{1/2}dx^{i}dx^{i}
+Z_2^{1/2}Z_{6}^{-1/2}V_{0}^{1/2}ds^{2}_{K3},\nn\\
 &e^{2\Phi}=g_0^2 Z_2^{1/2}Z_{6}^{-3/2}, \nn\\
 &C^{(3)}=(Z_2g_0)^{-1}dx^0\wedge dx^4\wedge dx^5,\nn\\
 &C^{(7)}=(Z_6g_0)^{-1}dx^0\wedge dx^4\wedge dx^5\wedge ({\rm vol}(K3)).
 \nn
\end{align}
Here, we use the harmonic function $Z_2$ and $Z_6$ of the form
\begin{align}
& Z_2=1-\frac{|r_2|}{r}, \qquad r_2
         =-\frac{(2\pi)^4g_0N\ls^5}{2V_0},\label{z2}\\
& Z_6=1+\frac{r_6}{r},  \qquad r_6=\frac{g_0N\ls}{2}.\label{z6}
\end{align}
In the above equations, $ds^2_{K3}$ is a Ricci flat metric on K3 of unit
volume, and ${\rm vol}(K3)$ is the volume form of the K3, which is
4-form.  We also used here the constant $\ls$ called ``string
length''. The string length characterizes length scale of the stringy
effects. If one want to see smaller things than $\ls$ or higher energy
things than $1/\ls$, he should use string theory rather than
supergravity.

In the above solution, there are two continuous parameter (integration
constant) $g_0,\ V_0$ and one discrete parameter $N$.  The $g_0$ is the
string coupling constant at far from the branes. The $V_0$ is the volume
of the K3 at far from the branes.

Let us consider the parameter region where the classical supergravity is
valid at least far from the brane. First, to suppress the loop
contribution of the string theory, $g_0$ should be small. We also
suppose $V_0$ is very smaller than the stringy volume 
$V_*:=(2\pi\ls)^4$ in order to avoid
the stringy effects.  Then, we can see from (\ref{z2}) and (\ref{z6})
that $|r_2|/r_6=V_*/V_0$ is very smaller than $1$, so we can conclude
$|r_2|\ll r_6$ when we suppose the supergravity is valid at infinity.
This means that if we observe this system far from the brane,
we cannot see the effect of $D2^*$ charge.

\begin{figure}
\begin{center}
  \mbox{\includegraphics[width=5cm]{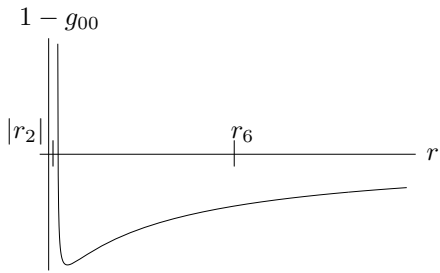}
   \put(4,42){$r$}
   \put(-150,93){$1-g_{00}$}
   \put(-70,50){$r_6$}
   \put(-70,42){$|$}
   \put(-155,50){$|r_2|$}
   \put(-138.5,42){$|$}
  }
\end{center}
\caption{The potential of {\dd} system. At $r=|r_2|$ the potential
 diverge. Near the singularity, there is repulsive force, and At far
 from the singularity, there is attractive force.}
 \label{potential}
\end{figure}

Now, we consider the singularity. The point $r=|r_2|$ is a naked
singularity. This singularity is ``repulsive'' which mean there are
repulsive force near the singularity. Therefore, this singularity is
called ``repulson singularity''. In figure \ref{potential}, we show
the gravitational potential.

Let us remark the validity of the supergravity near the singularity.
The singular point is $r=|r_2|=\frac12\left(\frac{V_*}{V_0}\right)g_0
N$. The validity of the supergravity at infinity requires both
$\left(\frac{V_*}{V_0}\right)$ and $g_0$ small.  If $N$ is not so large,
$|r_2|\ll \ls$ and supergravity or geometrical consideration is no more
valid. The region of repulsive force is the same order as $|r_2|$.
Therefore, If we intend to view the repulson singularity by supergravity
point of view, we should set $N$ very large in order to $|r_2|\gg
\ls$. However, even if we set $N$ very large, the compactification of
IIA supergravity is no more valid, because the volume of the K3
$V:=V_0Z_2/Z_6$ becomes arbitrary small near the
singularity. Especially, there is the
point $V=V_*=(2\pi \ls)^4$, which is called ``{\enh} radius'' in the
following context. At this point, extra massless fields appear as we see
later.

In the following subsection, we investigate what happens near the
singularity.

\subsection{Probe analysis}
In order to see the nature of the singularity, we use some object as a
probe and move it close to singularity.  A neutral object cannot reach
the singularity because of the repulsive gravity force. Then, how about
a charged object?  Especially, when we construct {\dd} system, we gather
$N$ of single {\dd} at a point.  Actually, {\dd}'s are BPS even if they
are separate, there are no force between them. 
For these reasons, we use a {\dd} as the
probe.  Do not confuse the source {\dd} system, and the probe {\dd}.
The former is composed of N {\dd}'s and makes the background.  The latter
is a single {\dd} used to study the geometry made by the source {\dd}
system.

Let us consider the D-brane dynamics on some nontrivial background.
For example, the action of a particle in the electromagnetic field
is
\begin{align*}
 S=-\int dt \;m\sqrt{-g}+\int eA^{(1)},
\end{align*}
where $g$ is induced metric, $A^{(1)}$ is pull-back the electromagnetic
field to world-line, $e$ is the charge of the particle, and $m$ is the
mass of the particle. On the analogy of a particle, a $p$-brane action
on $(p+1)$-form field $C^{(p+1)}$ background can be written as
follows. First, change the integration region to the $(p+1)$-dimensional
world volume. Next, change the length of the world line to the volume of
the world volume. Finally, change the gauge field to $(p+1)$-form which
is naturally integrated on the $(p+1)$-dimensional world volume. The
result is written as
\begin{align*}
 S=-\int d^{p+1}\xi T_p\sqrt{-{\rm det}g}+\int \mu C^{(p+1)}, 
\end{align*}
where $T_p$ is the brane tension, and $\mu$ is the brane charge.

In the case of the {\dd}, the action is a little complicated because of
the following reason.
\begin{itemize}
 \item The D-brane tension depend on the value of dilaton $\Phi$,
      the tension depend on $r$. When we integrate in the K3, the
      tension is also depend on K3 volume.
 \item As mentioned before, {\dd} system
     has negative D2 charge because of the curvature.
 \item Related above, the curvature of K3 has negative contribution
     to the tension.
\end{itemize}

As a result, if we integrate the K3 part, the action becomes
\begin{align}
 S=-\int_{M}d^3\xi e^{-\Phi(r)}(\mu_6 V(r)-\mu_2)\sqrt{-\det g}
+\mu_6\int C^{(7)}-\mu_2 \int C^{(3)},\label{probe-action1}
\end{align}
where $V(r)=V_0Z_2/Z_6$ as defined before, $\mu_p=(2\pi)^{-p}\ls^{-p+1}$
are D$p$-brane charges.

Now, we fix the reparametrization symmetry and assume some anzats.
\begin{itemize}
 \item Static gauge. $x^0=\xi^0,\ x^4=\xi^1,\ x^5=\xi^2$
 \item The prove move only to $r$-direction. We also assume $r$ is
       depend only on $\xi^0$.
 \item We consider the slow motion, and expand by the velocity
       $\dot{r}:=\del r/\del \xi^0$.
\end{itemize}
Then, the action (\ref{probe-action1}) can be rewritten as
\begin{align}
 &S=\int d\xi^1 d\xi^2 \int d\xi^0\left[(\text{const})
 +\frac{\mu_6 Z_6}{2g_0}(V(r)-V_*)\dot r^2+\dots
\right].\label{v-exp}
\end{align}

The $0$-th order term of the $\dot r$ expansion of integrand of action
is the potential term. In the action (\ref{v-exp}), the potential is
constant and independent of $r$.  Thus, if the probe is not move ($\dot
r=0$) then there is no force, and this system is stable. This is
consistent with the fact that this system (the source {\dd} system
and the probe {\dd}) is BPS if the probe stops.

\begin{figure}
\begin{center}
 \mbox{\includegraphics[width=8cm]{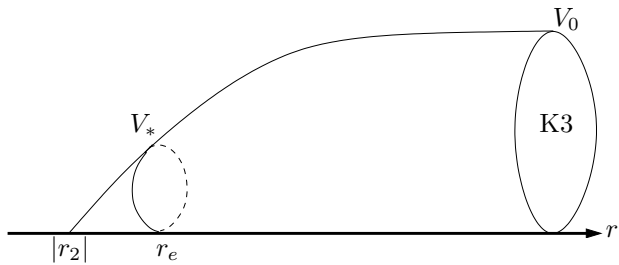}
  \put(0,0){$r$}
  \put(-170,-7){$r_e$}
  \put(-210,-7){$|r_2|$}
  \put(-25,40){K3}
  \put(-20,80){$V_0$}
  \put(-180,40){$V_*$}
 }
\end{center}
\caption{The image of the $r$ dependence of the volume of the K3 $V(r)$.
At large $r$, the volume is $V_0$ and we assume it is large. At $r=r_e$,
the volume becomes $V_*$, where the stringy effect is relevant.
At $r=|r_2|$, $V(r)=0$.}\label{fig-volume}
\end{figure}

The second term is the leading term of the kinetic energy. We can see
the tension depends on $r$, and when the volume of 
the K3 becomes the stringy volume: $V(r)=V_*$, the tension vanishes.
When the probe is far from the source branes, the volume
 of K3 becomes $V(r)\cong V_0 \gg V_*$.
If the probe approaches to the singularity and $r$ decreases, at the
point $V(r)=V_* \Leftrightarrow r=r_e:=\frac{2V_0}{V_0-V_*}|r_2|$ , the
probe becomes tensionless, and moreover, when $r<r_e$ the tension
becomes negative. This critical point $r_e$ is called ``{\enh} radius''.
The $r$ dependence of the volume of the K3 is shown in figure
\ref{fig-volume}.

Negative tension is rather pathological. It seems that near the {\enh}
radius, the classical supergravity is no more valid. Actually, when a
nonperturbative object like a D-brane become light, even the
perturbative string is no more valid. In the next section, we collect
some observations, and mention the interpretation proposed by Johnson,
Peet and Polchinski.

\section{Gauge symmetry enhancement and {\dd} shell} \label{sec-gauge}

\subsection{type IIA string theory on K3 as a gauge Higgs system}
In this section, we consider the 6-dimensional gauge theory obtained by
the compactification of type IIA string theory on K3, and for a while, let
us consider the background without {\dd} branes.

First, we look at two U(1) gauge field of the following. One is obtained
directly from ten dimensional $C^{(1)}$ by $C^{(1)}_m$ where
$m=0,1,\dots,5$. The other $\tilde C^{(1)}$ is obtained from
ten dimensional 5-form $C^{(5)}$(electro-magnetic dual of 3-form
$C^{(3)}$) by the relation $C^{(5)}=\tilde C^{(1)}\wedge {\rm vol}(K3)$,
where vol$(K3)$ is the volume form of the $K3$. Note that a {\dd} has a
magnetic charge $(1,-1)$ of $(C^{(1)},\tilde C^{(1)})$.

Next, we have a scalar field $H=V-V_*$, where $V$ is the volume of
the $K3$. The vacuum expectation value of $H$ is $V_0-V_*$.
Note that $H$ is a moduli, which means the vacuum expectation value
of $H$ takes continuous values, and $\vev H$ parameterizes
the vacuum.

Finally, there are a particles of {\ddp}. {\ddp} is the D4-brane wrapped
on K3, and there appear the negative D0-brane charge for the same reason
as the {\dd} case. We show the {\ddp} configuration in table
\ref{D4-D0}. We can see that it is actually a particle in 6-dimensions
from the table \ref{D4-D0}.

\begin{table}[h]
\begin{center}
  \begin{tabular}{|c|c|c|c|c|c|c||c|c|c|c|}\hline
    &0 &1 &2 &3 &4 &5 &6 &7 &8 &9 \\ \hline
    &\multicolumn{6}{|c||}{$\Rb^{5,1}$} &\multicolumn{4}{|c|}{K3} \\
    \hline D4 &\oo &\xx &\xx &\xx &\xx &\xx &\oo &\oo &\oo &\oo \\
    \hline D0${}^*$ &\oo &\xx &\xx &\xx &\xx &\xx &\xx &\xx &\xx &\xx \\
    \hline
\end{tabular}
\end{center}
 \caption{{\ddp} system}
 \label{D4-D0}
\end{table}

The action of the {\ddp} can be obtained by the same way as the {\dd}
case. The result is
\begin{align*}
 S=-\int d\xi^0 \frac{\mu_4}{g_0}(V_0-V_*)\sqrt{-g_{00}}.
\end{align*}
We can read the mass of the {\ddp} as $\frac{\mu_4}{g_0}(V_0-V_*)$ ,
which is proportional to the $\vev{H}$. If $V_0$ is not much larger than
$V_*$, then, {\ddp} particles is light, and easily pair created or pair
annihilated. Therefore, we should treat {\ddp} particles as
fields. Note that this field is electrically charged $(-1,1)$ under the
$C^{(1)},\tilde C^{(1)}$ so the field is complex.  Moreover, we assume
this field is a vector field. We denote this field as $W_{m}^{-}$ and
its complex conjugate $W_{m}^{+}$.

\begin{table}[htbp]
\begin{center}
  \begin{tabular}{|c|c|c|}\hline
 origin & field & interpretation \\ \hline\hline
 10-d 1-form & $C^{(1)}$ & photon \\ \hline
 10-d 5-form & $\tilde C^{(1)}$ & another photon \\ \hline
 volume of K3 & $H$ & massless Higgs field \\ \hline
 {\ddp} & $W_{m}^{\pm}$ & W-boson \\ \hline
\end{tabular}
\end{center}
\caption{The fields that we observe in 6-dimensional gauge theory
obtained from type IIA theory compactified on K3. They
 are interpreted as fields in a Higgs mechanism.}
\label{actors}
\end{table}

We summarize these actors in the table \ref{actors}. This shows that the
6-dimensional system includes the $U(2)$ gauge theory with massless
adjoint Higgs. The $U(2)$ gauge field is
\begin{align*}
 A_{m}=\left(
\begin{array}{cc}
C^{(1)}_m & W^+_m \\
W^-_m & \tilde C^{(1)}_m
\end{array}\right).
\end{align*}
Actually, the charge of the {\ddp} is consistent with this picture.

What is the role of the {\dd} in this picture? It has magnetic charges
of the photon. Hence, we can conclude the {\dd} is the
'tHooft-Polyakov monopole in this gauge Higgs system. Near the center
of the classical 'tHooft-Polyakov monopole, the value of Higgs becomes
small and the W-boson mass becomes small.

When $H=0$ the W-bosons become massless, and the gauge symmetry is
enhanced from $U(1)^2$ to $U(2)$. At the {\enh} radius of {\dd} system,
this gauge symmetry enhancement occurs. The name ``{\enh}'' is from this
gauge symmetry {\em enhancement}.

\subsection{{\Enh} as a shell}
Let us go back to the {\dd} system and consider the problem that
the tension of the probe becomes negative. In the field theory,
this is not a problem. In that case, only the mass square
is appear in the theory, which mean $m_{W}^2\propto \vev{H^2}$. 

In the probe theory, we can make a trick and avoid a negative mass as
follows. We can rewrite the probe action (\ref{probe-action1}) outside
the {\enh} radius as
\begin{align}
 S=-\int d^3\xi e^{-\Phi(r)}\mu_6\sqrt{-(V(r)-V_*)^2\det g_{ab}}+\dots.
\label{probe-action2}
\end{align}
If we use this form also inside the {\enh} radius, the mass does not
become negative, but another problem occur. Inside the {\enh} radius,
the potential term of (\ref{probe-action2}) is not constant. It becomes
\begin{align}
 S=\int d^3\xi\left[\frac{2\mu_6}{g_0}(Z_6^{-1}(r)V_0-Z_2^{-1}(r)V_*)
+(\text{higher order in }\dot r )\right].
\end{align}
The potential is flat when $r>r_e$. On the other hand, when $r<r_e$,
there are repulsive force. Therefore, {\dd} probe cannot stop at a point
inside the {\enh} radius and the nearest point it can stop is the {\enh}
radius.

\begin{figure}
\begin{center}
\mbox{\includegraphics[width=5cm]{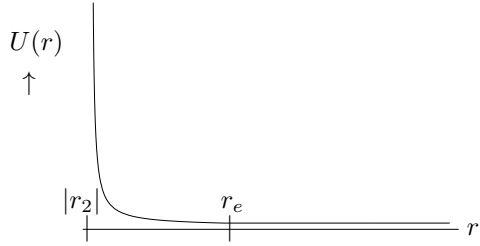}
  \put(3,0){$r$}
  \put(-170,70){$U(r)$}
  \put(-165,55){$\uparrow$}
  \put(-90,10){$r_e$}
  \put(-88,0){$|$}
  \put(-150,10){$|r_2|$}
  \put(-142,0){$|$}
}
\end{center}
\caption{The graph of potential term in (\ref{probe-action2}).
 $U(r)=-\frac{2\mu_6}{g_0}(Z_6^{-1}(r)V_0-Z_2^{-1}(r)V_*)$.
 If $r>r_e$, the potential is flat. If $r<r_e$, there are repulsive force.}
\end{figure}

\begin{figure}[htbp]
\begin{center}
  \mbox{\includegraphics[width=10cm]{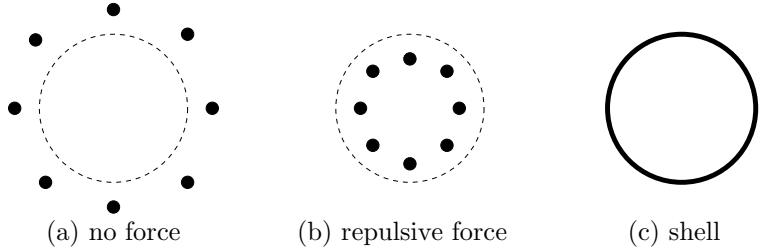}
   \put(-270,-10){(a) no force}
   \put(-175,-10){(b) repulsive force}
   \put(-50,-10){(c) shell}
}\end{center}
\caption{To make singularity by gathering the {\dd}. The dashed circle
 expresses the {\enh} radius. If {\dd}'s are put as (a),
the configuration is BPS and there is no force. On the other hand,
if {\dd}'s are more close each other and put as (b),
there are repulsive force and the configuration is unstable. So the
closest stable configuration is the shell as (c).}
\label{fig-shell}
\end{figure}

Now, let us consider how to construct the {\dd} system and repulson
singularity by gathering $N$ {\dd}'s. Let us observe one of the $N$ {\dd}'s, and others as sources
of the background field, that is we separate {\dd}'s to a probe brane
and source branes. This is a kind of mean field approximation.  The
prove {\dd} can be stops anywhere when it is apart enough from the
source branes. Then, we move slowly the whole {\dd}'s to a point to make
the singularity.  But when they are closer each other than the {\enh}
radius, there are repulsive force and the configuration is unstable.  In
the result, the closest stable configuration is one in which the {\dd}'s
form a $S^2$ shell at the {\enh} radius. The image of this observation
is shown in figure \ref{fig-shell}

In the interior of the shell, it seems that the metric is flat because
there are no source in the interior the shell. Consequently, there are
no singularity in this configuration. This mechanism of resolution of
singularity is called ``{\enh} mechanism''.

There is still another problem. We naively mention that {\dd}'s are form
the shell. However, if we treat {\dd}'s as finite number of thin object,
they cannot form a whole shell. More concretely, there are SO(3)
symmetry in the original {\dd} system, but we cannot realize a SO(3)
invariant configuration by putting finite number of points on the $S^2$.
Therefore, on the {\enh} shell, the branes are no more thin, but they
seem to melt, mixed to each other and become a higher dimensional
object. We review this picture from another point of view in the next
section.

\section{{\Enh} and fuzzy sphere}
\footnote{This section was omitted in the talk.}
\label{sec-fuzzy}

In order to see how the {\dd} branes form a $S^2$, let us go to a dual
picture --- the D3-D5 system. The D3-D5 system is, as shown in table
\ref{D3-D5}, composed of 2 D5-branes and
$N$ D3-branes attached to both D5-branes.  The two D5-branes correspond
to the curvature of K3 in the {\dd} system, and $N$ D3-branes correspond
to $N$ {\dd} branes in the {\dd} system. Therefore the {\enh} can be
seen in the D3-D5 system as D3-brane shell. Let us see how the D3-branes
form a shell. 

\begin{figure}
\begin{center}
  \mbox{\includegraphics[width=4cm]{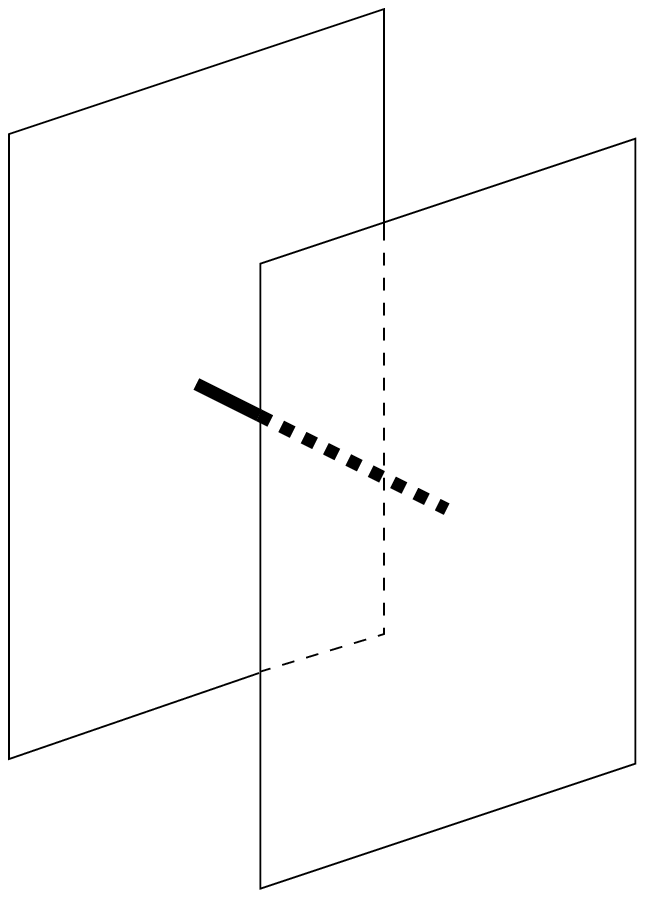}}
 \hspace{1cm}
 \mbox{\includegraphics[width=4cm]{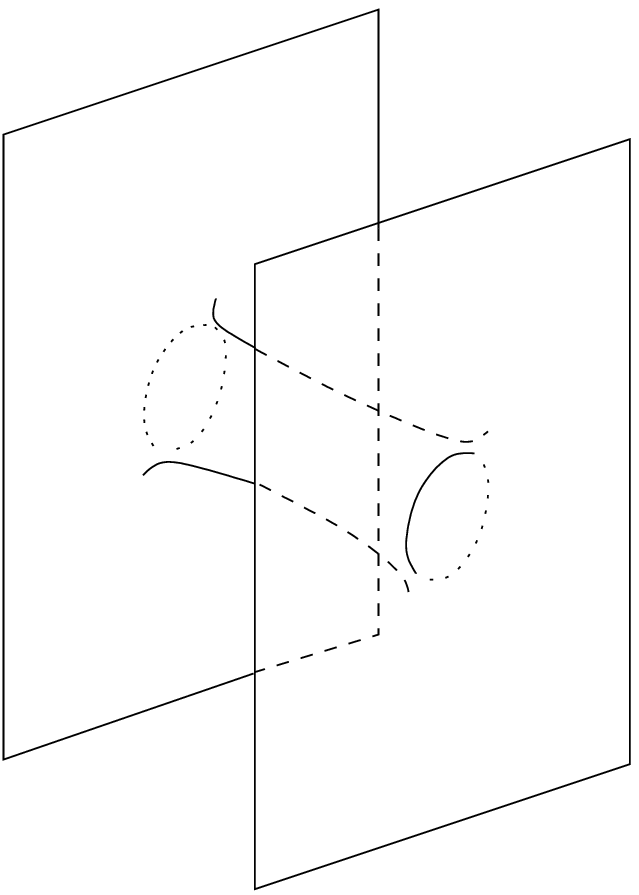}}
\end{center}
\caption{The image of the D3-D5 system. The D3-branes are attached to
 two D5-branes. From the point of view of the gauge theory on D5-branes,
 the D3-branes are smeared and become a tube. The smallest radius of the
 throat is to be {\enh} radius\cite{Johnson:2000qt,Johnson:2001bm}.}
 \label{double-funnel}
\end{figure}

From the point of view of the gauge theory on the D5-branes,
the D3-branes are the magnetic monopole --- 'tHooft Polyakov monopole.
classical solution of 'tHooft Polyakov monopole is not a point like
thin object, but a fat object. In brane picture, this is a tube like
configuration as shown in figure \ref{double-funnel}.

\begin{table}[h]
\begin{center}
  \begin{tabular}{|c|c|c|c|c|c|c||c|c|c|c|}\hline
    &0 &1 &2 &3 &4 &5 &6 &7 &8 &9 \\ \hline
    \hline D3 &\oo &\xx &\xx &\xx &\oo &\oo &\oo &\xx &\xx &\xx \\
    \hline D5 &\oo &\oo &\oo &\oo &\oo &\oo &\xx &\xx &\xx &\xx \\
    \hline
\end{tabular}
\end{center}
 \label{D3-D5}
 \caption{D3-D5 system}
\end{table}

Let us consider this system from the point of view of the gauge theory
on the $N$ D3-branes. The theory on the D3-branes is the super symmetric
U($N$) gauge theory. It includes the gauge field $A_{\mu},\ \mu=0,4,5,6$
and adjoint scalar field ($N\times N$ hermitian matrix)
$X^I,I=1,2,3,7,8,9$. The matrices $X^I$ are interpreted as the coordinate
of the position of the D3-branes. We use the following anzats here.
\begin{itemize}
 \item Only three of the scalar $X^1,X^2,X^3$ have nontrivial value.
       The other field: the gauge field and $X^7,X^8,X^9$ is set to be 0.
 \item $X^1,X^2,X^3$ depend only on one of the space-like coordinate
       $\sigma$ of the world volume.
\end{itemize}

Since we want BPS solutions, we should consider the Bogomolnyi equation
in stead of the equation of motion.  If a configuration satisfies
Bogomolnyi equation, that configuration also satisfies the equation of
motion.

In the above anzats, Bogomolnyi equation becomes
\begin{align}
 \del_{\sigma}X^i=\frac i2\varepsilon^{ijk}[X^j,X^k].
\label{Nahm}
\end{align}

There are trivial solutions in which the $X^{i}$'s are independent of
$\sigma$.
\begin{align*}
 \del_{\sigma}X^i=0,\quad [X^i,X^j]=0,\qquad (i,j=1,2,3).
\end{align*}
In this case, $X^i,\ i=1,2,3$ can be diagonalized at the same time.
\begin{align}
 X^i={\rm diag}(x_1^i,x_2^i,\dots,x_N^i).
\label{trivial-solution}
\end{align}
This shows that the position of the $a$-th D3-brane is
$(x^1_a,x^2_a,x^3_a)$ . So this solution is $N$ of thin D3-branes,
which means that the position of each brane is sharply determined.
The solution (\ref{trivial-solution}) is not the solution we need.
In this solution, there are no D5-branes, and not the dual of original
{\dd} system.

\begin{figure}
\begin{center}
  \mbox{\includegraphics[width=4cm]{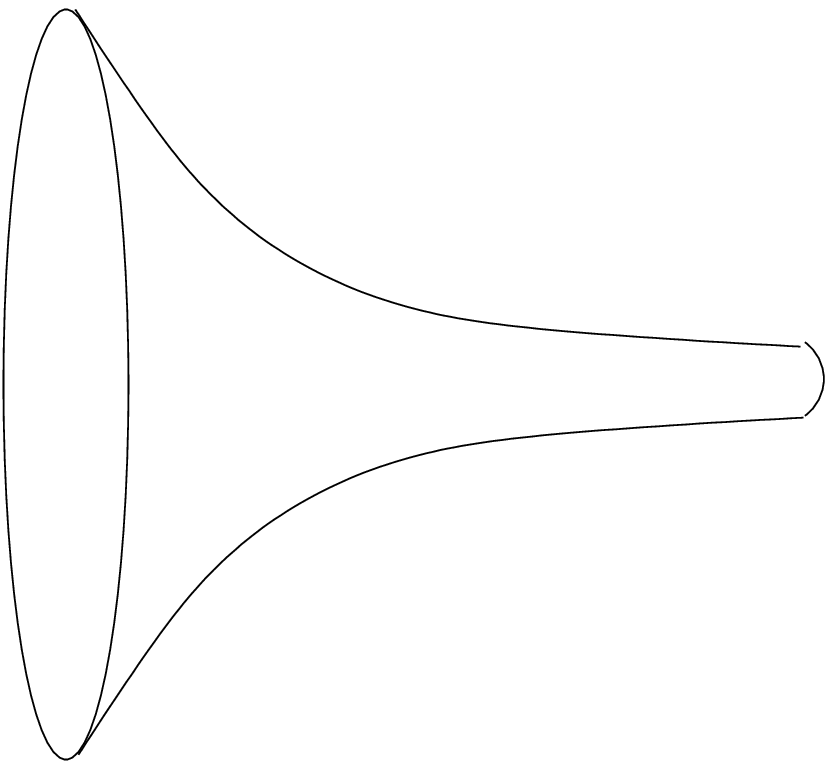}
     \put(-30,70){$\rightarrow \sigma$}
     \put(-80,80){$\uparrow R(\sigma)$}
  }\hspace{1cm}
  \mbox{\includegraphics[width=4cm]{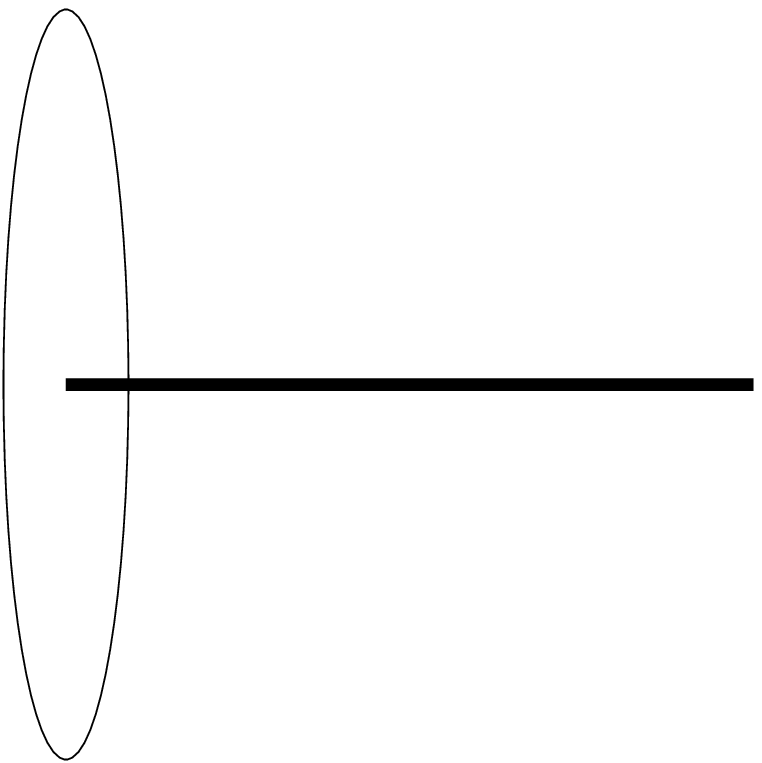}
     \put(-30,60){D3-brane}
     \put(-90,80){D5-brane}
  }
\end{center}
\caption{The image of the funnel solution. The section of constant
 $\sigma$ is a $S^2$ of radius $R(\sigma)\propto \sigma^{-1}$.
This solution can be interpreted as blurred picture of 
$N$ D3-branes ending on a D5-brane. Actually it is known that this
solution has a D5-brane charge.}\label{funnel}
\end{figure}

We show here another solution.
The shape of this solution is like the figure \ref{funnel}, and
called ``funnel solution''\cite{Constable:2000ac}.
 In that solution, the position of each
branes are blurred and has SO(3) symmetry. These are necessary
properties of the {\enh}, and we regard this solution as {\enh}.
The solution is
\begin{align*}
 &X^i(\sigma)=-\frac{1}{\sigma}\alpha^{i},
\end{align*}
where $\alpha^{i}$ is $N$-dimensional representation matrix of
SU(2) which satisfy
\begin{align*}
 [\alpha^i,\alpha^j]=i\varepsilon^{ijk}\alpha^k.
\end{align*}
In this solution, from the Casimir operator of $N$-dimensional
representation, the following relation is satisfied.
\begin{align*}
 (X^1)^2+(X^2)^2+(X^3)^2=(R(\sigma))^2\cdot 1_{N},
\qquad R(\sigma)^2=\frac{1}{\sigma^2}(N^2-1),
\end{align*}
where $1_{N}$ is the $N\times N$ identity matrix.
As a result, if we consider the constant $\sigma$ section,
$(X^1,X^2,X^3)$ is formally a $S^2$ of radius $R(\sigma)$.
This space is called ``fuzzy sphere''\cite{Madore:1992bw}.

This geometry can be interpreted as the figure \ref{funnel}. 
The section of constant $\sigma$ is a ``$S^2$'' of radius $R(\sigma)$
$R(\sigma)\propto \sigma^{-1}$ and divergent at $\sigma=0$. This can be
interpreted as there are a D5-brane at $\sigma=0$. So this is a picture
of a D5-brane and $N$ D3-branes end on the D5-brane from the point of 
view of the gauge theory on D3-branes. Actually, it is shown that
this solution has a D5-brane charge\cite{Myers:1999ps}.

We obtained a solution which contains a D5-brane and $N$ D3-branes.
However, strictly speaking, this is not the solution we need.  In this
solution, there is only one D5-brane and the gauge group on D5-brane is
U(1), and this solution is Dirac monopole in view of the gauge theory on
the D5-brane. The solution we actually need is shown in figure
\ref{double-funnel}. In this case, there are two D5-branes and the gauge
group on D5-brane is U(2). There are adjoint Higgs fields on the brane
gauge theory, and the gauge group is broken to U(1)${}^2$. Moreover, the
D3-brane is a 'tHooft-Polyakov monopole from the viewpoint of
D5-brane gauge theory. To obtain this type of solution
is a future problem.

\section{Conclusion}

In this paper, we review the {\enh} mechanism --- a mechanism of
resolution of repulson singularity of {\dd} system. In this mechanism,
the {\dd} makes a shell at the {\enh}, which is in exterior the
singularity.
Since there is no source in the interior of the shell,
the inside of the shell is flat and nonsingular. This
shell turns out to be a fuzzy sphere.

We mention the recent progress about {\enh} here. In \cite{Jarv:2000zv},
other types of 3 dimensional gauge theory appear by introducing some
kinds of orientifolds. The investigations from the view point of gauge
theory on the brane are also done in \cite{Evans:2000ct,%
Bertolini:2001dk,Frau:2001gk,Petrini:2001fk,Billo:2001vg,
Merlatti:2001gd}. In these papers, {\enh} and the gauge theory on
``fractional D-branes'' are studied. Fractional D-branes are wrapped
D-branes on vanishing cycles of some (for example orbifold) singularity
and their D-brane charge is actually fractional. {\enh}s appear also in
these fractional D-brane system.  In
\cite{Johnson:2001us,Constable:2001fe} {\enh} and the black hole
thermodynamics are investigated. In \cite{Maeda:2001si}, the supergravity
solution of {\enh} shell and its stability are studied. In
\cite{Johnson:2001wm} more on consistency of {\enh} are corrected and
applied to some general case.

\subsection*{Acknowledgement}

I would like to thank the participants of the associated seminars,
especially, Yoshifumi Hyakutake, Akihiro Ishibashi, Hideo Kodama, Makoto
Natsuume, Jiro Soda, Masa-aki Sakagami for usefull discussion.

The work of the author is supported in part by the JSPS Research
Fellowships for Young Scientists.

\end{document}